# The Limit and Its Ground. QBism, Scientific Realism, and the Born Rule

Harald A. Wiltsche, Linköping University

QBists insist that theirs is a realist position while denying that the quantum formalism represents a mind-independent world. The insistence is hard to place. The realism debate is throughout a dispute about represented content, so a theory that denies standing in the representation relation appears not to take up a position within that debate at all. This paper examines the most developed attempt to make good on the QBist claim—David Glick's perspectival normative realism—and argues that it fails instructively. Glick seeks the ground of the Born rule's objectivity among states of affairs and construes an agent's perspective as a location within an already-constituted world; finding no state of affairs fit to bind every agent at once, he must book the ground as brute. Phenomenology supplies the third thing his picture lacks. A *horizon* is not a position within the world but the condition under which a world is given. Read as an agent's weighting—her curvature—of such a horizon, the quantum state is neither a defective description nor a brute posit, and the Born rule articulates the invariant lawfulness by which any sense confers a curvature at all.

Keywords: scientific realism, QBism, phenomenology, transcendental philosophy, horizon, Born rule

## Introduction

The debate over scientific realism is usually staged as a dispute about how theories relate to a mind-independent world: whether their claims are candidates for truth, and whether empirical success gives us reason to believe them. QBism unsettles this staging before the argument even starts. According to its proponents, the quantum formalism does not describe an observer-independent reality at all; it is, in their words, "a decision theory concerning the actions and consequences of any agent wishing to better navigate the world" (DeBrota, Fuchs, and Schack 2024, 14). If quantum theory is a manual for acting rather than a picture of what there is, the question of its realism cannot be posed in the usual terms—and yet QBists insist that theirs is a realist position all the same. This chapter examines the most sophisticated attempt to make good on that insistence—David Glick's perspectival normative realism—and argues that it fails for reasons that point beyond QBism itself, toward a different, phenomenological conception of what a perspective is.

The rejection of standard forms of scientific realism is the consequence of QBism's staunch anti-representationalism, i.e., the scepticism about conceiving of quantum theory as a depiction of a mind-independent world. This anti-representationalist stance is not an idiosyncratic interpretive choice: according to the proponents of QBism, it is forced by specific technical results, including a dissolution of the EPR nonlocality worry (Fuchs, Mermin, and Schack 2014; Fuchs 2017a) and the Pusey-Barrett-Rudolph no-go theorem for ψ-epistemic interpretations of the quantum state (Pusey, Barrett, and Rudolph 2012; Berghofer 2024, sec. 3.2). What replaces representation, on the QBist view, is a decision-theoretic self-description: quantum theory as "a tool decision-making agents are advised to adopt in light of the peculiar uncertainties we find in our world" (Fuchs 2023, 82), or, more compactly, as "an empirically motivated addition to Bayesian decision theory" (DeBrota, Fuchs, and Schack 2024, 1).

Given this self-description, QBism has repeatedly been charged with being "nothing but instrumentalism in disguise" (Hagar 2003, 772)—a natural worry, but one that does not survive contact with what "instrumentalism" actually names. On an older, Nagel-style reading, theories are not truth-apt at all: like a hammer or a drill, they are tools to be used well or badly, not descriptions capable of being true or false (Nagel 1961, ch. 6, sec. III). A QBist state assignment does not fit this mold—unlike a hammer, it is true or false: true when it is the assignment an agent's own information obliges her to make, false when it is not, though never true or false of a mind-independent fact (Healey 2023, sec. 2.2). On the more familiar contemporary reading, associated with positions such as van Fraassen's constructive empiricism, instrumentalism is an epistemic stance: unobservable claims are neither believed nor disbelieved (McMichael 1985, 258). But this reading presupposes exactly what QBism denies—that the formalism's claims are truth-apt representations of a mind-independent world about which one could remain agnostic in the first place.

QBism thus fits neither sense of the instrumentalist label. Nor do QBists themselves accept it. They insist, constantly and explicitly, that theirs is a realist position (Fuchs 2017b; Fuchs, in Crease and Sares 2021)—just one that draws on a sense of "realism" quite unlike the ones standardly on offer in philosophy of science. Making sense of that insistence is where the argument of this chapter actually begins.

**QBism for Real**

QBists' realist self-descriptions have been remarkably consistent, even as their precise formulation has evolved. The earliest and best-known is Fuchs's "participatory realism," coined explicitly against the charge of instrumentalism: rather than "relinquishing the idea of reality," QBism holds that "reality is more than any third-person perspective can capture" (Fuchs 2017b, 113). This is meant as a positive ontological claim, not a retreat: the universe, on this view, is no "block universe," fixed, out there, and ready for us to discover; there is no "view from nowhere," and, correspondingly, no complete, third-personal description of it available to any agent (ibid.). The same conviction underwrites QBism's account of measurement as an "act of creation [or] births," where what quantum theory is fundamentally about is "how those births or acts of creation impact the agent's expectations for other such births" (ibid, 122)—a formulation meant to secure reality not despite quantum indeterminacy but by way of it: measurement outcomes are treated as the growing edge of an unfinished world, not as failures to disclose an already-complete one.

This same ambition resurfaces, considerably sharpened, in Fuchs's more recent coinage "normative structural realism," introduced in a 2021 interview with Crease and Sares. Pressed on what, precisely, is real for the QBist if neither quantum states, measurement operators, nor unitary evolutions are, Fuchs locates the realism not in any of the formalism's terms but in the relation between them—above all the Born rule:

> None of the terms in the Born rule are ontic in themselves, but the *rule* plays an ontic role [...]. If you're looking for things in the theory that strive to be statements about reality independent of a particular agent, these have got to be among them. This is why I call these aspects a *normative structural realism*. (Fuchs in Crease and Sares 2021, 555; my emphasis)

Structural realism about physical theories is not itself a new position—it holds, roughly, that what survives theory-change, and hence what a theory is really tracking, is not the world's object-level ontology but its relational structure. Fuchs borrows this vocabulary but detaches it from its usual anchor: the structures in question are neither ontic (describing objective states of affairs) nor epistemic (describing objective knowledge of them), but *normative*—constraints on how any agent ought to relate her own probability assignments to one another. In the same interview Fuchs makes the underlying intuition explicit: while the "particular evaluations or elements we each use [the] methods [of quantum mechanics] upon […] are personal, […] the method itself is transpersonal, with a third-person flavor" (ibid., 546). It is this third-person flavour that, according to Fuchs, gives QBism its realist bite. We come back to this issue in the next section.

The picture that emerges is this: Fuchs admits that QBism has "made much of its living by re-classifying" (ibid., 550) pieces of the quantum formalism as subjective. This tendency notwithstanding, however, Fuchs emphasizes that QBism has "never strayed from what Einstein called 'the program of the real'"—that is, the idea that, first, talking about "the real" is one of the essential tasks of physics, but that, second, physics earns the right to speak of "the real" not by fiat but by giving a working, empirically fruitful account of what is objective and what is merely a function of an observer's situation (ibid.). As Fuchs describes the ambition elsewhere, the "very goal of QBist research has been to distill a statement about the character of the world from the fact that the gambling agents within it should use the quantum formalism"; QBism's task, so understood, is "to reverse engineer from the formalism to a characterization of an ontology" (Fuchs 2023, 97f.).

We are now at a point where we can see clearly what I consider to be the main challenge for QBism. Fuchs wants an ontology, but what he in fact gets is subjectivism plus a rule that tells subjects what they ought to do. The trouble, then, is not QBism's sincerity about wanting to be realist, but the shape of the claim itself: *an ought does not obviously entail an is*. That any agent who fails to follow the Born rule is vulnerable to a Dutch book tells us something about the costs of incoherence; it does not, by itself, tell us anything about the world the Born rule is a rule for coping with. The question, in other words, is whether a normative structure can secure realism about anything at all, and if so, how.

Note that this is a slightly different worry from other well-known complaints about QBism's realist ambitions. For example, Harvey Brown grants that QBism does not deny the existence of an observer-independent reality—but presses on what that concession actually amounts to. If the nature of that reality is left so unspecified that nothing can be said about it beyond its bare existence, Brown argues, the position becomes indistinguishable in practice from a view that denies it outright. He puts the comparison directly: "For Berkeley, the nature of God's action in creating living minds is a mysterious affair. That's the way it is with God. I fail to see how the action of the external world on human agents in QBism is any less mysterious" (Brown 2019, 81).

Ryckman, writing from a considerably more sympathetic starting point, reaches a related, but distinct verdict: the move from QBism's "bedrock empiricism"—the assumption that experience always belongs to a particular agent—to any realist conclusion about "that objective world out there" leaves us, in his own words, in "a frustrating void" (Ryckman 2023, 193) because it remains unclear how the ascent from subjective experience to claims about objective reality is supposed to work in detail. Both worries deserve to be taken seriously on their own terms, but neither is quite the difficulty pressed here. Brown's target is the bare claim that an external world exists at all; Ryckman's is the leap from an empiricist starting point to any realist conclusion whatsoever. The question this chapter pursues is narrower and more specific: not whether enough can be said about an external world, or whether empiricism can support realism in general, but whether a rule addressed to what agents ought to do could, even in principle, secure realism about anything—and, if it can, how.

**Glick's Perspectival Normative Realism**

The problem that the previous section ended with is precisely the difficulty that David Glick's "perspectival normative realism" is designed to solve, and it is the most developed attempt in the literature to do so on QBism's own terms (Glick 2021). Glick begins by showing that QBism fits uneasily with what he calls standard scientific realism—typically broken down, following presentations such as Psillos (1999) and Chakravartty (2025, section 1.2), into a metaphysical component (a mind- and theory-independent reality exists), a semantic component (theories are to be read literally), and an epistemic component (theoretical claims constitute knowledge of the world). Regarding the metaphysical dimension, QBism is not obviously at odds with it: QBists speak of an external world that delivers "unpredictable kicks" (Fuchs 2017b). The real problem lies with the semantic and epistemic dimensions. Both are standardly cashed out in terms of representation: theories are to be read literally, and taken to constitute approximately true knowledge, of how segments of empirical reality actually are. Yet, as we have seen, QBism denies that quantum mechanics does either of these things—which is why it sits fundamentally at odds with standard scientific realism.

According to Glick, the most serious candidate for aligning QBism with a form of scientific realism is to view it as a limiting case of selective structural realism—a view on which almost nothing in the formalism corresponds to reality except the Born rule. Recent QBist work lends this reading some support: the reformulation of the Born rule in the SIC-based "Essential Representation" eliminates the quantum state altogether and recasts the rule as a relation among an agent's subjective probabilities (DeBrota, Fuchs and Stacey 2020), while Fuchs's suggestion that the rule "correlates with something that one might want to call real" (Fuchs 2017b, 6), together with his remarks on Hilbert-space dimension as marking a "universal capacity" of matter (Fuchs 2010), gestures toward exactly the kind of residual, formalism-independent purchase on reality that selective realism requires. Glick nonetheless judges the reading a failure, and a failure on realism's own terms rather than QBism's. A position on which only a bare relational rule survives is, first, deeply uninformative about what is real: it tells us that *something* corresponds to the Born rule without telling us what. It leaves the world it is meant to be realist about "unspeakable" (Timpson 2008) or "ineffable" (Brown 2019)—verdicts that sit oddly with any position still claiming the name of realism. And it cannot do the one piece of explanatory work selective structural realism is normally invoked to do: answer the no-miracles argument. If the rule is purely relational, and if what it relates are agents' subjective credences rather than anything answerable to a shared world, nothing in the picture explains why coherent betting by this rule should be reliable rather than merely

self-consistent—the theory's success is left exactly as mysterious as it was before the realist gloss was added.

What distinguishes Glick from critics who treat QBism's divergence from standard realism as simply a failure on QBism's part is that he takes the divergence as diagnostic rather than damning. Whatever realism QBism can earn, Glick concludes, it will not be earned by claiming that the formalism, however minimally, mirrors the world—which leaves open whether an entirely different conception of realism is needed.

In his quest to find such an entirely different conception of realism, Glick's first step is to reconstruct QBism as *radical* perspectivalism, as opposed to conservative perspectivalism (Glick 2021, 10). The latter (Glick reads Rovelli and Dieks in this way) relativizes states and properties to agents but still delivers a description of external reality; the perspectival facts supervene on the God's-eye facts plus the agent's situation. QBism goes further since it abandons the descriptive ambition altogether. Its perspectival facts therefore fail to supervene even on a complete God's-eye description of reality together with a full specification of the agent's situation. Fixing all the objective facts and locating the agent among them still does not determine which quantum state she will assign; agents in the very same situation may licitly assign different states. Quantum theory, on this reading, applies only at the level of agents embedded in the world in a particular way.

This is also what allows Glick to accommodate QBism's participatory rhetoric. Talk of measurement as an "act of creation" (Fuchs 2017b, 122) is not a metaphysical proposal about external reality but a claim to be assessed at the level of the situated agent—much as, on a familiar treatment of secondary qualities, a tomato's being red is a genuine feature of the perceiver's situation rather than an illusion, without being a property the world possesses from no point of view (Glick 2021, 11). Measurement outcomes are, in this sense, features of agents; they may ultimately be grounded in facts about external reality, but the connection is not straightforward. The denial of a God's-eye description is thus, in Glick's words, "a claim about the limits of quantum theory, not the impossibility of metaphysics by other means" (ibid., 12). The QBist remains free to believe in a mind-independent reality; what she may not do is read that belief off the formalism.

With the appropriate sense of perspectivalism clarified, Glick proceeds to what he considers to be the decisive point: according to QBism, quantum theory is fundamentally *normative, not descriptive*. The Born rule tells an agent what she has reason to do—how to set her credences—not what the world is like. Yet crucially, realism about a normative theory need not carry an underlying metaphysics—here Glick leans on Scanlon's moral realism, on which metaphysics matters only insofar as it is needed to account for the reason-giving force of the theory's prescriptions. What must then be supplied is a *grounding* for the Born rule's normative authority. Since the inductive grounding available to Healey (2017) is closed to the QBist—there are no objective frequencies to appeal to, and the Born rule is relational—what remains is the de Finetti/Dutch-book strategy: the Born rule is an "empirically extended coherence" constraint (Fuchs and Schack 2013, 1702), formally like the probability axioms but requiring more than logic and mathematics—it binds "only in a world relevantly similar to ours" (Glick 2021, 15). However, as Glick readily admits, *why* our world makes the Born rule the correct coherence constraint is something the QBist presently cannot say: it is left as a brute feature of reality, "nature's whisper" (Fuchs 2017b, 6). We shall come back to this point.

Glick's conclusion: QBism meets the realist demands appropriate to a normative theory—objective coherence constraints answer the "no right answer" worry; the Dutch-book argument supplies reason-giving force—while the demand for causal-mechanistic explanation is dismissed as question-begging. It is question-begging to expect—as, for instance, McQueen (2017) does—substantive explanations of quantum phenomena because offering such explanations is simply not what quantum theory, according to QBism, is designed to do. QBism thus marks a *limit* of scientific realism—exactly as the title of Glick's paper advertises. It is as realist as a radically perspectival, fundamentally normative theory can be—and no more.

**The Limits and Their Ground**

Glick's is the most serious attempt to make good on QBism's realist self-description, and it is worth being clear about what it gets right before pressing on where it runs out. Two achievements should be preserved. The first is negative: Glick shows convincingly that the standard realist apparatus—literal construal, approximate truth, models latching onto worldly structure—cannot be the frame in which QBism's realism is assessed, and that reading QBism as a limiting case of selective realism yields a position too thin to be informative. The second is constructive: he sees that if quantum theory is fundamentally normative, then the realism appropriate to it is a realism about the bindingness of its prescriptions, and that such a realism owes an account of what grounds that bindingness. Both points survive what follows. Indeed, the criticism depends on them: it is because Glick asks the right question—what makes the Born rule an objective constraint?—that the shape of his failure to answer it is instructive.

For the answer he arrives at is that there is no answer to be had. The Born rule is an objective coherence constraint for agents in a world relevantly similar to ours; why our world is like that is "nature's whisper," a brute feature of reality with which the QBist must rest content. My claim in what follows is not that this concession is premature, as though more ingenuity within Glick's framework would recover the missing ground. It is that the framework is built so as to place the ground out of reach—and that the difficulties I canvass below are so many symptoms of one structural feature: Glick seeks the ground of the Born rule in the register of a state of affairs, and construes the agent's perspective as a location within a world constituted independently of it.

The first problem becomes apparent once we acknowledge that there are two commitments in Glick's account pulling in opposite directions. Glick needs the Born rule to bind every agent: this is what his realism comes to, once description has been given up—not that the formalism is true of the world, but that its prescriptions are objectively in force for anyone who uses the theory. Yet he also needs perspectivalism to be radical, so that no God's-eye description together with a specification of the agent's situation fixes her assignments. But these commitments are irreconcilable. A constraint binding all agents alike is naturally understood as a constraint answerable to something those agents hold in common; radical perspectivalism is precisely the denial that there is any such common structure to answer to. The objectivity of the norm and the radicality of the perspective cannot both be had for free, and Glick owes an account of how they consist together.

The account he offers is the phrase "a world relevantly similar to ours" (Glick 2021, 15). The Born rule, he says, is an objective coherence constraint not everywhere but in worlds of a certain kind—ours among them—and this is meant to locate the norm's objectivity without reinstating a description of how those worlds are. But the phrase cannot bear the weight. To individuate worlds and rank them by similarity is to help oneself to exactly the descriptive resources Glick—and, by extension, QBism—has disavowed: one cannot say which worlds are relevantly like ours without saying, in some register, what ours is like. And the gloss he supplies when pressed—a world "in which quantum theory provides a good guide for agents in it" (ibid., 15)—circles back on itself, since a world in which the Born rule binds just is what was to be explained. What began as the ground of the norm's objectivity turns out to name the fact that it holds.

Glick does not disguise where this leaves him. Asked what it is about our world that makes the Born rule the correct coherence constraint, he grants that "there is very little the QBist can say" (ibid.), and settles for treating it as a "brute fact" (ibid.)—Fuchs's "nature's whisper" (Fuchs 2017b, 6). It would be a misreading to take this as a candid admission of unfinished business, a gap that further work might close. The gap is not contingent. Having denied himself any structure the agents share, Glick has denied himself in advance anything that could ground a constraint binding them all; the brute fact is not where the explanation has not yet reached but where, given the framework's commitments, no explanation can be posed. The ground of the norm's objectivity is not merely unstated but placed beyond statement. This is the first and clearest instance of the pattern announced above: the ground is sought in the register of a state of affairs—some feature of the world in virtue of which the rule holds—and, none being available, is booked as brute.

The second problem concerns the move Glick makes once the first has done its work. If no feature of the world grounds the Born rule, perhaps none is required: quantum theory is normative rather than descriptive, and a normative theory answers to a different kind of realism—one that asks not whether its claims are true of the world but whether its prescriptions are genuinely binding. The Born rule, so understood, is a coherence constraint, akin to the axioms of subjective probability: an agent who violates it is not mistaken about some fact but *incoherent*, exposed in principle to a sure loss. This is meant to convert the missing ground from a liability into a non-issue.

But it changes the subject rather than settling it. The demand that drove the search for a ground was not imported from outside; it was Glick's own. His realism, recall, is a realism about *bindingness*—not that the formalism is true of the world, but that the Born rule is objectively in force—and in force, as he is careful to say, not everywhere but in a world "relevantly similar to ours" (Glick 2021, 15). That restriction is not idle: it is what makes the position a realism about *our* world rather than a truth of pure reason. And it is exactly what a coherence constraint cannot underwrite. For coherence is world-unrestricted. The axioms of subjective probability bind an agent in any world whatever; a body of credences can be perfectly coherent and systematically false, since Dutch-book immunity guarantees only that the credences hang together, not that they answer to anything beyond themselves. So the recasting faces a scissors: Either the Born rule's bindingness is nothing but coherence—in which case the qualifier "relevantly similar to ours" does no work, and the realism about our world evaporates into a truth of rationality that would hold whatever the world were like. Or its bindingness is world-restricted after all, holding here and not elsewhere—in which case coherence is not what grounds it, and the ground we were promised is still missing. Recast as a coherence constraint, the Born rule is insulated from the difficulty only by being severed from the very world-restriction that made it a realist thesis in the first place. The debt is not discharged but declared uncollectible.

Glick's own model for the manoeuvre shows what it would take to make it work—and what is missing. As mentioned, he appeals to Scanlon, for whom realism about right and wrong need carry no moral metaphysics: the point of moral judgment is practical, and metaphysics matters only so far as it is needed to account for the reason-giving force of the judgments (Scanlon 1998). But Scanlon can decouple normative realism from metaphysics because he has somewhere else to anchor the bindingness of reasons—a substantive conception of the agent as a reasoning being, responsive to considerations that count for and against acting.[1] The reasons are binding in virtue of what the agent is. Glick borrows the shape of the move while leaving the corresponding place empty: his agent is a bare bearer of credences, with no structure in virtue of which the Born rule could be binding for her.

It may seem the place can be filled more cheaply. One need not credit the QBist agent with Scanlon's apparatus of self-governance to find her a ground; it is enough, the thought runs, that she is a coherent Bayesian—responsive, minimally, to the norms of probabilistic reason, and exposed to a sure loss if she flouts them. The bindingness of the Born rule would then be nothing more exotic than the bindingness of coherence itself, and no feature of the world, and no structure in the agent beyond her rationality, need be invoked. But this underdescribes what the Born rule asks. Coherence, in the Dutch-book sense, constrains the *form* of a credence function—that it be additive, normalised, consistent across the bets it licenses; it does not fix the particular relations the Born rule imposes among the assignments an agent makes for distinct measurements. That the two come apart is not my charge against QBism but QBism's own: the reconstruction programme running from Fuchs's work onward is built on the thesis that the Born rule is no theorem of probabilistic coherence but an *addition* to it—a normative constraint the world, in Fuchs's phrase, "whispers" to agents who reason within it. By the QBist's own accounting, then, the coherent Bayesian is left underdetermined with respect to the very rule at issue; the norms the defender invokes are precisely the ones the Born rule was said to exceed. The appeal to Bayesian rationality does not supply the missing ground but returns us to where we began. For what the coherent agent lacks, and what the rule's specific content demands, is some tie to the particular character of the world she measures in—and to name that character is either to re-enter the descriptive register QBism has disowned or to book it, once again, as brute. The Bayesian route is not an alternative to the difficulty of the previous section; it is that difficulty in another guise. Neither the analogy with Scanlon nor the retreat to bare coherence, then, relieves the demand for a ground; each relocates it. If the objectivity of the Born rule is not to be found in the world, and is not to be a brute fact, it must lie in something about the constitution of the agent for whom the theory is a guide—not a further fact about her psychology, but a condition on her having a world to be guided in at all. That is a question neither Glick's framework nor QBism is equipped to pose—and it is the question the remainder of this chapter takes up.

---

[1] Scanlon's rational agent is, at root, a reasoning creature: one capable of recognising, weighing, and being moved by reasons (Scanlon 1998, ch. 1). This capacity for reasons-responsiveness is what makes an agent a bearer of judgment-sensitive attitudes—beliefs, intentions, fears—attitudes an ideally rational agent would hold or abandon according to her own judgment of whether there is sufficient reason for them. Because these attitudes answer to the agent's judgment, they are hers in the sense relevant to appraisal: their attributability to her is what grounds her responsibility for them (ch. 6). Within the contractualist structure this reasoning capacity is further specified as a capacity for self-governance: the agent's actions and attitudes depend, at crucial junctures, on how reasons seem to her as a conscious individual, and she is moved by the aim of living with others on terms none, similarly motivated, could reasonably reject (chs. 4–5). It is this structure—reasons-responsiveness, judgment-sensitivity, self-governance—that lets Scanlon ground the bindingness of reasons in what the agent is, without further moral metaphysics. Glick's agent—or QBism's, for that matter—has no counterpart to any of it.

The pattern shows itself not only at the foundation, where the ground goes missing, but in the local devices Glick uses to make the position livable. Consider first the analogy with secondary qualities, offered to explain how measurement outcomes can be "features of agents" without lapsing into idealism (Glick 2021, 11-12). To be red, on the familiar model, is to be disposed to appear red to a suitable observer under suitable conditions; redness makes essential reference to a perceiver without being a mere projection. Glick proposes to treat outcomes likewise: real for the situated agent, yet not features of the world from no point of view. But the analogy carries a commitment he cannot honour. What keeps colour a respectable secondary quality, rather than a free-floating appearance, is a describable primary base—surface microstructure, reflectance, illumination—in virtue of which the object is disposed to appear as it does. Strip that base away and "appears red" grounds nothing; retain it and one has reinstated exactly the describable, observer-independent reality QBism denies. The outcome is a dilemma. Either the analogy keeps its primary base, and the representationalist picture returns through the back door, or it drops the base to stay faithful to QBism, and the outcomes become the bare projections the charge of idealism alleged. Both horns rest on a single premise[2]—that what grounds an appearance is either some describable state of affairs or else nothing—and it is this premise, not any incidental feature of the colour case, that the analogy cannot survive. Glick's "the connection is not straightforward" (ibid., 11) names the gap precisely at the point where it needs to be closed.

The coherence machinery strains in the same direction (Glick 2021, sec. 5.1). A Dutch book has force only if its bets can be settled—and here the QBist can grant more than one might expect. Each bet, taken by itself, is settled by the agent's having the determinate experience she has; Fuchs asks for nothing more, and nothing more is needed to close a single wager. But a sure loss is not a feature of any single settlement. A Dutch book is a package of bets placed together, and its force lies in their joint structure: that these outcomes are alternatives on one and the same measurement, mutually exclusive and exhausting the space, so that a credence function is defined over them at all. That is what Glick's outcomes cannot supply. He has told us they are "elements of subjective experience" and "as such completely unconstrained" (ibid., 14n19). Private episodes so described may each be determinate for the agent and still fail to compose an algebra: nothing in them fixes that this outcome and that one are alternatives on a single measurement rather than two unrelated happenings, and without that there is nothing over which incoherence could be read off. The sure-loss verdict presupposes a world already articulated into determinate alternatives over which betting is so much as defined. The apparatus meant to ground the norm without leaning on the world turns out to lean on it after all—not on the world as an inventory of states of affairs, but on a world constituted to a particular level of articulation. And here the grounding runs the wrong way for Glick: coherence over bets is not the floor beneath the norm but itself rests on a prior achievement, the constitution of a field in which there is anything determinate to bet on.

[2] One might resist the premise with the relationalist, holding that colour is grounded not in a primary base but in the standing relation between object and perceiver. Yet this opens no third way. For the relation to keep colour respectable it must have a worldly relatum with a describable character—a reflectance, a surface—on pain of relating the perceiver to nothing; and that relatum is the base once more. A relation whose worldly pole had no describable character would no longer be the familiar secondary-quality model, but the very reconception still owed.

Both devices send Glick back into the register he had hoped to leave behind. Colour needs a describable base; the Dutch book needs a settled field. In each case the account, at the moment it must make contact with the world, either becomes the description it disowns or gestures at a ground it cannot state. This is the pattern of the first problem recurring in the machinery, and it is the second half of the diagnosis coming into view: the difficulty is not merely that Glick seeks the ground in the register of a state of affairs, but that he construes the agent's perspective as a position within a world already constituted—so that whenever he reaches for what a perspective rests on, only states of affairs are there to be found.

Return, in closing, to the taxonomy with which Glick's reconstruction began. Radical perspectivalism was introduced as a thesis about the world: unlike the conservative perspectivalism of Rovelli and Dieks, on which the perspectival facts supervene on the God's-eye facts together with the agent's situation, QBism's perspectival facts were said to supervene on no such base. Yet when the point is pressed, Glick draws it back. In a footnote he concedes that measurement outcomes may "ultimately supervene on the God's eye facts and the agent's situation—it's just that the former are largely unknowable according to QBism" (ibid., 10fn15) and relocates the distinctiveness of QBism to the failure of supervenience on the *accessible* perspective-free facts. The metaphysical thesis has quietly become an epistemic one. What was a claim about what the world does or does not fix is now a claim about what the agent can or cannot know—and conservative and radical perspectivalism, so understood, differ only over the reach of ignorance. And the concession costs more than the taxonomy. If outcomes supervene, after all, on a God's-eye base, then the "brute fact" of the earlier sections was never groundlessness but ignorance—a ground that exists and fixes the Born rule, merely out of reach; QBism's anti-representationalism shrinks to modesty about an inaccessible base, and the perspective-free reality it set out to disown returns as a noumenon it cannot see.

This slide is not carelessness; it is forced. The supervenience question that frames the whole taxonomy—do the perspectival facts supervene on the God's-eye base together with the agent's situation?—can be posed at all only if the agent's situation is itself specifiable in God's-eye terms, as one more region of an already-constituted world. Both horns accept this much; they agree that a perspective is a position within such a world, and quarrel only over whether the position fixes the facts. But once perspective is construed as a location, there is nowhere for the ground of the Born rule to lie except among the states of affairs that furnish the world—which is exactly why, at every point where a ground was wanted, only states of affairs were there to be found, and the norm had in the end to be bruted. Glick oscillates between a metaphysical and an epistemic reading of "radical" because perspective-as-location leaves him no third thing to say: either the facts float free of the base, courting the solipsism he means to avoid, or they are fixed by a base he cannot reach, reinstating the God's-eye description he means to reject. The way forward is not to choose a horn but to refuse the picture that generates both. What if a perspective were not a place within the world but the horizon through which a world is given to an agent at all—not a further fact, apt to supervene or fail to, but the condition under which anything comes to count as a fact? Developing that thought is the work of the rest of this chapter.

**A Way Out—the Phenomenological Notion of Horizon**

Before building my own argument, it is worth marking where the argument stands, since the positive proposal answers to a problem that has now taken definitive shape. QBism began from a refusal: the quantum state does not describe an empirical target system in the external world but expresses an agent's expectations, and the Born rule does not report a law of nature but prescribes how those expectations must hang together. This is what made QBism more than instrumentalism and less than representation—and what left it owing an account of its own realism. For if the formalism describes nothing, the sense in which quantum theory answers to the world must lie not in the truth of its claims but in the bindingness of its rule; and a realism about bindingness owes an account of what grounds it.

Glick's perspectival normative realism is the most serious philosophical attempt to pay the debt: the Born rule, he proposes, is an objective coherence constraint, binding not for this or that agent but for any agent in a world relevantly similar to ours. But we saw where this runs out. Pressed for what it is about *our world* that makes the rule the correct constraint, Glick can answer only with Fuchs's "nature's whisper," and books the ground as brute—and the brute fact is not a gap further work might close but the mark of a structural choice. Having construed the agent's perspective as a location within a world already laid out, he left the ground of the rule nowhere to lie among the states of affairs that furnish that world: and none being fit to bind every agent at once, it had in the end to be bruted. The moral was not that Glick chose the wrong state of affairs but that he sought the ground in the wrong register. What is needed is not a better-specified world but a different construal of perspective: not a position within the world, but the condition under which there is a world to take up a position in. Phenomenology has a name for perspective so understood—the *horizon*—and a worked-out account of how it structures the givenness of objects. The rest of the chapter puts that account to work, building it up through three ordinary illustrations before turning to physics: an open field of what a situation leaves undetermined, a personal weighting laid over that field, and an impersonal rule connecting the two—what will be called, respectively, the horizon, its curvature, and the invariant relation between sense and curvature. The thought that the phenomenological notion of horizon could do this work is not new. De la Tremblaye (2020) already reads QBism as implicitly phenomenological and puts the horizon at the center of her reading by stating that "the perceptual horizon parallels the QBist quantum state" (255). What follows builds on that instinct, though not, in the end, on her way of cashing it out.

Consider first an ordinary case, before any physics. I look across the street at a house. What meets my eye is a façade—one face, at one distance, in this light; but what I *see*, in the full sense, is a *house*, and a house is not a façade. The far side I do not see is co-meant with the side I do see; so are the rooms behind the wall, the depth the frontage promises, the standing of a thing that has an inside. None of this is inferred after the fact from the coloured surface before me. It is intended along with that surface, as the penumbra of further experience the house prescribes: what I would find were I to cross, walk around, go in. Husserl calls this fringe of co-intended but unfulfilled possibilities the *horizon* of the experience, and marks its two reaches (Husserl 1973, §8). There is the *inner* horizon—the object's own unshown aspects and the ways it would answer to my dealings with it—and the *outer* horizon—the co-present surroundings in which it stands, the street running on past the edge of my attention, the town, the world, given not as a further object but as an unregarded background without limit. To perceive the house is to intend it *as* a house, and to intend it as a house is already to have this horizon in play. Strip it away and nothing is left given: not a house seen more austerely, but a flat patch of colour with nothing behind or around it—which is to say, no thing at all.

Notice what carries the object's identity through this perceptual episode. As I cross the street the façade slides from view and a side swings into it; the appearances succeed one another without rest. Yet I do not take myself to meet a new object at each step. I take the changing appearances as *one* house showing itself variously, and I can do so because through their flux I hold fast to invariants—the lawful way the sides give onto one another as I move, the constancies that persist while the perspectival show varies. Objectivity, even here, is not won by climbing out of perspective to a view from nowhere. It is won *within* perspective, by holding constant what the perspective alters. The point will matter shortly; for now, it is enough that the horizon is what makes an identical object available across a manifold of appearances, none of which is the object itself.

Three aspects must be emphasized at this point. First, although there is a problematic tendency in the literature to present the phenomenological notion of horizon as if it was an exclusive feature of visual perception, horizonal givenness is a necessary feature of *any kind of experience*. Reaching into my bag without looking, I close my hand on my keys. What is actually felt is little—a cold shaft, a jag of teeth, the curve of the ring—but what is given is the keys: the blades I have not yet touched, the heft they will have as I lift them, the fit they promise in a lock across town. I tell them from the loose change beside them through a horizon of further touches I anticipate rather than perform. Nothing here is seen; the horizon is wholly tactile—and yet it is the same structure, the felt given ringed by the co-meant unfelt, without which my hand would close on a meaningless hardness and not on my keys. Or take a mathematical case. Consider what is given when I grasp the sequence of natural numbers. I do not, and could not, run through them; what stands before me is at most a few—one, two, three—together with an "and so on" that carries the sense of the whole. That "and so on" is a horizon: to grasp the numbers *as* the numbers is to intend, beyond every figure actually reached, a determinate manner of going on—each with its unique successor, none of them the last. The infinity of the series is nowhere given, yet it is co-meant in every instance of it. Here nothing is sensed but a few inert marks; the horizon that lets them be the numbers is not sensuous but categorial—founded on those marks, yet nowhere among them, the field of what the sense "natural number" prescribes.

Second, since horizonal givenness is typically introduced through concrete examples in which individual subjects undergo horizonally structured experience, one could get away with the impression that the phenomenological notion of horizon describes the psychology of such subjects—a set of empirical facts about how creatures like us happen to process what is set before them. That impression must be dispelled, and much of what follows depends on dispelling it. The concrete experiences of concrete subjects are indeed horizonally structured; but the horizon is not one of the facts that make up such an experience, on a footing with the firing of the subject's nerves or the beliefs she carries to the scene. The horizon is rather that in which any such fact is given as a fact at all. Being the condition of facts and not a fact among them, it is no one's in particular—not a private field each subject carries about, as she carries her nerves and her beliefs, but the one structure within which any subject's experience is articulated at all, however first-personally it must be shown. It is, in that precise sense, impersonal (Berghofer and Wiltsche forthcoming).

This distinction—between a matter of fact and the condition under which anything is a fact at all—is not a mannerism of phenomenology's idiom; it can be seen directly, and without taking any side in the metaphysics of tense, in the case of time. Take a clock: this pointer at this graduation is a state of affairs, and the physics of time treats such states exhaustively—durations, simultaneities, coordinate times. Yet a reading is legible as a reading only for a consciousness that already holds the moment just past and anticipates the moment not yet arrived, so that a present can stand out from both—what Husserl calls retention and protention. This lived temporality is not one further fact beside the clock's; it is what lets anything, the clock's own reading included, count as a temporal fact in the first place.

To classify the horizon, likewise, among the empirical furniture of the mind is not merely false but category-mistaken: it mistakes for one of the given things the very light in which anything is given at all, and so must presuppose, in the very act of psychologistic reduction, the horizonal givenness it means to reduce away. This is why no result in the psychology of perception could confirm the notion or refute it, and why nothing in what follows hangs on the contingent make-up of human cognition—which is not to place the structure beyond revision, only beyond the reach of direct empirical test. To summarize: The horizon is not a state of affairs that the sciences of mind might one day inventory; it is a transcendental condition—not, to say it at once, in any sense that trades on a reality beyond experience, but in the precise sense of that under which anything is given to experience as the object it is.[3]

Third, the horizon so far described might be taken for a flat structure—a bare inventory of the possibilities an experience leaves open, each on a par with the rest. It is nothing of the kind. To intend the house is not merely to co-intend a manifold of further experiences but to hold them in a definite order of anticipation—and to hold them as answers to courses I might take. Were I to round the corner, a back would all but certainly show, as would a roof overhead and an interior behind the wall: these are anticipations so heavily weighted that their default would be not a surprise but a collapse of the sense "house" itself. Were I to look for a way down from the hall, a cellar might or might not be found—admissible, but held more loosely. Were I to open any given door, nothing in the sense "house" would speak for or against the room beyond being painted yellow: neither anticipated nor excluded, simply left open. And no course I could take would show me a façade with nothing behind it, or a house that dissolved as I came round it—not entertained, but positively excluded. The horizon is curved: what it holds open is graded, from the all-but-certain through the merely admissible to the ruled-out, and this grading belongs to it as intended rather than being laid over it after the fact. The grading, moreover, is not a weighting of free-standing possibilities but of what would show, course by course; the weights attach to explorations, not to properties standing on their own. A possibility that is nobody's possibility of exploring is, phenomenologically, no possibility at all.[4] Husserl, borrowing from the probability theorist Johannes von Kries, calls this weighted field the *Spielraum*—the "room to play" every experience keeps ahead of itself, which is not the horizon's openness but its curvature, the differential loading of its possibilities by the sense under which the object is intended (Lobo 2019, 519–20; von Kries 1886).

---

[3] Transcendental structures are revisable, but not by anything they make possible—much as the axioms of a geometry cannot be overturned by any proposition expressible within it. No proposition of Euclidean geometry refutes the parallel postulate, since every theorem already presupposes it: the postulate is immune from within. It is not thereby beyond revision. One may consistently deny it, and for the space-time of general relativity a non-Euclidean geometry is the more adequate, not Euclid's. So too with the structures at issue here: unfalsifiable by anything they constitute, yet open to being overtaken when a domain comes to be constituted more adequately through other means.

Two features of this loading will bear weight later; the first is the kind of magnitude it is. The weight a continuation carries is *intensive*, not extensive. An extensive magnitude is assembled by summation—so many units of length, so many members of a class—and a frequency is its paradigm, a ratio of counts. The curvature of a horizon is not assembled that way. That a back would all but certainly show were I to round the corner is not the quotient of a census of houses; it is the degree to which the sense under which I intend the thing *speaks for* that continuation, here and now, given all the situation has laid down. The weight is *founded*, in Husserl's sense—resting on what counts in favour of the anticipation—rather than tallied from a record of cases. This is not to make it innate. The sense that does the weighting is itself sedimented from a life of experience; what is denied is only that the weight is a stored frequency. A founded weight is a present degree of warrant conferred by a sense, not a ledger of how often. That is why the horizon would be curved just so for a subject who had kept no count at all: where a frequency needs a history to be a frequency *of*, a founded weight needs only a sense and a situation. And for most of the horizon there is nothing to count in any case, since the courses to which the weights attach are courses not taken—the door I never open gives onto nothing whatever, and the weight it carries is definite regardless.

---

[4] The indexing to what I might do should not be understood as making the horizon mine, in tension with the impersonality established above. What is indexed is not a repertoire this or that subject happens to command, but the form of horizonal givenness as such: horizons are articulated by courses of exploration, and it is that articulation, not anyone's inventory of capacities, to which the grading attaches.

The second feature is the curvature's objectivity, which is of the kind already met with the house—won within perspective, not beyond it—and untouched by what was just conceded about sedimentation. That a sense is a personal acquisition, laid down over a history of experience, does not make the curvature it confers a personal accident: whatever sense one has come to command confers upon the horizon the very curvature it would confer on anyone who commanded it. The relation of sense to weighting is an eidetic law, not a contingency of this or that formation. What is objective, then, is neither a subject's coming to command some sense—itself a personal, sedimented achievement—nor some curvature two subjects happen to share, but the invariant lawfulness with which any sense delivers its weighting.

With the horizon in hand—transcendental rather than psychological, impersonal, weighted rather than flat—we can return to the quantum state, and to the impasse it left us in. What defeated Glick was that he could locate a perspective only as a place within the world, so that the ground for the Born rule had nowhere to lie but among the states of affairs that world contains—and, none of them fit to bind every agent at once, it had to be bruted. The horizon is the third thing his picture could not supply: a perspective that is not a position in the world but the condition under which a world is given. Suppose, then, that the quantum state belongs to such a horizon—not as the horizon itself, but as the curvature an agent lays over it. On the QBist's own telling it already stands close: a bundle of an agent's expectations for what her measurements will yield, held in different degrees of expectancy, narrowing to a single outcome when a measurement is made and reforming, revised, for what comes next. Set that against the horizon built up above. The outcomes a measurement leaves open make up the field—an open structure, the same for any agent who faces that measurement; what she brings to it is a weighting, the curvature her sense confers as she takes up her situation. The field is one of outcomes for that measurement, not of properties the system has anyway: here as with the house, the weights attach to explorations and not to what stands on its own.

That is QBism's own insistence, reached from the side of givenness rather than from the formalism. The agent's sense, in the quantum case, is nothing occult: it is what she brings to the measurement—the system as she has prepared it, her prior dealings with systems of its kind, the information in light of which her expectations take the shape they do. The state is the form that curvature takes when it is written down. Read so, the state is not a *state of affairs* at all, defective or otherwise—and it is not the horizon either: it is the agent's curvature of the horizon. The outcomes the field holds open, the state their weighting made exact, the Born rule the rule by which that weighting is read. Nothing in this makes the horizon the agent's to dictate. Anticipation can be disappointed, and disappointment is not a failure of the structure but the way transcendence announces itself within it: a weighting that could never be overturned would not be the curvature of a horizon but a closed picture with nothing beyond it. Here Fuchs's unpredictable kicks belong—not a reality outside the account knocking at it, but the disappointment of an anticipation, and so already an event in the field they are said to disturb. The field, correspondingly, is not laid down by the agent's sense; it is what her sense runs up against, articulated in the meeting and not in advance of it.

This is where de la Tremblaye's equation needs amending, not abandoning: the horizon she rightly puts at the centre is never the state. Once field, curvature, and invariant are told apart, this also settles two difficulties French has pressed against her account: that a transcendental structure is confused with an agent's private web of expectations (2023, 200), and that a structure drawn from ordinary experience is asked to bear weight it was never fitted for in the quantum case (197). The first does not arise, since the horizon is never the state; the second is met by the observation, above, that horizonal givenness is no more perceptual than it is tactile or categorial.

But if the quantum state is an agent's own curvature—personal to her because the sense she brings to her situation is her own—has objectivity not simply been surrendered? Two agents may assign different states to the same system, and nothing here forbids it—this is the radical perspective Glick, following QBism's own commitments, insisted on, and the very point at which he despaired of a ground. The same point answers a cognate worry: perspectives, on view's like Massimi's, are markedly more flexible than horizons as characterized here, generating the familiar Hacking-style plurality about a single system (Massimi 2022)—different measurement choices directed at "the same" electron. But different measurement choices are, on this account, different situations, and different situations open different horizons: the plurality is not a cost to the notion but an instance of it. Indeed, objectivity, on the account presented here, was never going to lie in a curvature two agents share. It lies, as it did with the house, in the invariant lawfulness by which any sense confers its curvature at all—the eidetic relation that holds for whoever takes up the sense, whatever curvature results. The state is personal; the relation is not. And this is what the Born rule is. A curvature is always some agent's; the Born rule is not a curvature but the relation every curvature answers to—the constraint the same for every agent, whoever she is, which any personal state must honour. The lawfulness invoked here is fully general: it governs the house and the keys no less than the measurement. The Born rule is not that lawfulness as such, but what it comes to for a world constituted to the level of articulation at which quantum phenomena can show themselves. The generality secures that there is a constraint and that it is nobody's to legislate; the constitution of the domain makes it this one.

That the state is personal and the relation impersonal is already the solution to the first difficulty we pressed against Glick: objectivity and radical perspective, which would not cohere for him, cohere now that objectivity is lodged in the relation and perspectival radicality in the state. And the same relocation reaches further, unmaking the taxonomy with which his reconstruction began. Its organizing question was whether the perspectival facts supervene on the God's-eye base together with the agent's situation—conservative perspectivalism answering yes, radical perspectivalism no—and it was in wavering between a metaphysical and an epistemic reading of that "no" that Glick's position slid, because perspective-as-location left him no third thing to say. But the question cannot be put to a horizon. A horizon is not a perspectival fact that might supervene on the base or float free of it; it is that in whose light there is a base, and facts to supervene, at all. To ask whether it supervenes on the God's-eye facts is already to have miscast it as one more fact among them—the very miscasting this whole detour has been undoing. The oscillation Glick could not escape therefore does not arise. The state, read as the agent's curvature of a horizon, is neither fixed by a base beyond reach nor floating free of every base: it is grounded—but in the sense the agent brings, not in the God's-eye facts. And that sense is no region of the base on which anything might supervene; it is the way the agent takes up her situation, a condition of givenness and not one of the things given. The state is thus not a further item to be located among those facts, but a weighting of the very field in whose light they stand.

The local devices settle with it. The secondary-quality analogy wanted a describable base and the Dutch book a settled field—each, we found, driving Glick back into the register he meant to leave, and each reaching for the horizon without a name for it. The base beneath the appearance and the field over which betting is so much as defined are not further states of affairs the account must locate, but the constituted horizon within which appearances and wagers have a place.

**Two Modes of Representation**

The account just given relocates the ground of the Born rule; but it does not yet say what becomes of the realism that was in question. That question was never idle. QBists insist that they are realists, and this chapter began by taking the insistence seriously rather than treating it as a slip. What remains to be shown is that the horizon delivers a realism worth the name—and that it is the one QBism has been or at least should be reaching for.

Begin with why the insistence sounded so odd. Standard realism holds that theories are to be read literally and are approximately true of a mind- and theory-independent world; constructive empiricism holds that we may accept them without believing what they say about the unobservable; structural realism holds that what survives theory change is relational structure rather than object-level furniture. However far apart these positions stand, they share a conception of what representation is: a correspondence between a theory and an observer-independent state of affairs, its truth a matter of how well it tracks that domain. They differ over the reach of our access to it, not over the kind of relation at stake. And because the dispute between them is throughout a dispute about represented content—whether we may believe it, how much of it survives theory change, which parts of it latch on—a theory that denies standing in that relation does not take up a position within the debate so much as drop out of it: there is nothing left for the realist and the anti-realist to disagree about. QBism denies exactly that relation. Its insistence that it is nonetheless realist was bound to sound like a person insisting they are still in a room they have visibly walked out of.

Whether the fault lies with the person or with the walls is what Glick's attempt allows us to decide. He set out to place QBism among the realisms without disturbing the conception of representation the debate assumes, and every device he reached for failed in the same way. And the failures share a commitment he never puts on the table: that the only things a claim can answer to are observer-independent states of affairs. The secondary-quality analogy foundered on it: what grounds an appearance is either some describable state of affairs or else nothing. The Dutch book required a field over which betting is so much as defined; that field being no state of affairs, the account had to lean on a world it had undertaken not to lean on. And the taxonomy offered the perspectival facts a God's-eye base to supervene on, or no base at all. This is the same commitment the standard realism debate makes, in a different vocabulary. There it appears as a thesis about representation—to represent is to correspond to a state of affairs; here as a thesis about grounding—to be grounded is to be grounded in a state of affairs. Both are faces of a single restriction on what there is to answer to. And that is why QBism cannot secure its realism by declining to enter the debate. The restriction is not a convention of the genre, to be left behind along with it; it governs any attempt to say what the quantum formalism answers to, which is why it defeated the one attempt that had conceded QBism's anti-representationalism from the outset. It must be denied outright—and denying it means exhibiting something that is answered to and is not a state of affairs.

The horizon is such a thing. It is not a state of affairs—not one of the things given, but that in whose light anything is given—and it is not nothing, since it is structured, weighted, and lawful. But then there is more than one kind of thing a formalism can answer to, and answering cannot mean the same in both cases. To answer to an observer-independent state of affairs is to match it: the description mirrors how things stand, and its truth is a matter of how well it tracks that domain. Call representation in this mode *thick*. To answer to a horizon cannot be to match it *in this sense*, since a horizon is not a way things stand. It is to articulate the constraints under which things can stand in some way at all—constraints that are real and non-arbitrary, yet do not exist independently of the situation they structure. Call this mode *thin*. What a thin representation yields is not a picture but an articulation—not of how things are, but of how they hang together; and it is therefore not a thick representation with its claims lowered, not a partial or approximate mirroring, since what it articulates is not a domain at all, and so not a domain dimly seen.[5]

This is what the QBist personal/impersonal distinction has been reaching for without a vocabulary to state it, and it makes sense of a formulation of Fuchs's that otherwise holds together by nerve alone: "None of the terms in the Born rule are ontic in themselves, but the *rule* plays an ontic role." (Fuchs, in Crease and Sares 2021, 555) The first clause is what the QBist arguments establish—no term of the formalism thickly represents anything. The second is what those arguments leave untouched, and what the horizon now allows us to state: the rule thinly represents the invariant. Seen in this way, it now also becomes clear that "normative structural realism" is not a compound of convenience. What we are talking about is normative because a curvature is something an agent can get wrong, the weighting her situation warrants being no guarantee of the weighting she adopts; structural because what the rule articulates is not any particular curvature but the form every curvature in its domain must share; and realism because that form, though not beyond history, is at no agent's disposal. It is also what it takes to remain within what Fuchs, following Einstein, calls the "program of the real"—the demand that physics earns its right to speak of the real by giving a working account of what is objective and what is a function of the observer's situation. The distinction between an agent's curvature and the form any curvature must have is such an account, and it locates the objectivity in the invariance rather than in anything's obtaining with nobody there.

[5] The distinction is drawn, and argued at greater length, in Malone and Wiltsche (forthcoming), in a neighbouring quantum-mechanical dispute: whether the entangled state of an observer-apparatus-system must be read as describing how that composite stands. The formulations here are adapted to the present argument. One feature of that account is worth carrying over. Thick representation is not simply an error: within the natural attitude—apparatus calibrated, couplings specified, pointer-readings compared—a state is assigned with a robust object-directed relation in mind and with full legitimacy at that level. The mistake is not to represent thickly, but to take that thickness for the last word on what the formalism does.

**Conclusion**

The question this chapter began with can now be answered. QBism's realism looked unplaceable not because it was confused but because the debate in which a place was sought for it recognizes one mode of representation where there are two. Read for thick representation, the quantum formalism says nothing whatever, and its practitioners' insistence on realism is bluster. Read for thin, it articulates the invariant structure of the horizon within which any agent's expectations take shape, and the insistence is exactly right.

Nothing in this displaces QBism's own self-description. Quantum theory remains a decision theory: the Born rule tells an agent what she ought to expect and what she may stake on it, and in that sense it guides action rather than describing a domain. What the foregoing shows is only that guiding action and representing are not alternatives. The rule can bind an agent because it articulates the structure of the situation in which she must decide.

A QBist might nonetheless want none of it. Part of the appeal of the position has always been its austerity: agents, their gambles, and a rule that keeps the gambles coherent. Set beside that, horizons and senses and curvatures and eidetic laws look like a great deal of machinery imported to close a gap the QBist was content to leave open. Why should a theory that has made a virtue of travelling light take on this much freight? My answer: Because nothing has been taken on. Nothing has been added to the inventory of the world: the horizon is not a further item in it, and the argument throughout has been that treating it as one is precisely the mistake. What has been added is a distinction— between what is given and that in whose light anything is given at all—and a distinction is not an entity. The structure was always in play. What was missing was any acknowledgement of it, and the cost of the omission is on record. It is why Glick, having sought the ground of the Born rule among states of affairs and found none fit to bind every agent at once, had in the end to book it as brute. And it is why QBists, whenever they try to say what their formalism tells us about the character of the world, must fall back on metaphor—kicks, whispers, nature's whisper again.

Yet what this chapter has not done should also be said plainly. It has relocated the ground of the Born rule and identified the mode of representation at work, but it has not given the reconstruction—the detailed account of how the rule arises as the invariant form of horizonal curvature, and of how such an account stands to the QBists' own derivation from coherence together with "only a little more." This is a debt of a different order from the one we found in Glick. His framework left the question unposable—nothing could have counted as an answer to it; here the question has a determinate form, and it is one the QBists' own reconstruction program is already pursuing in another idiom. Pursuing it from the phenomenological side is the work of a companion to this chapter.

**Acknowledgments:** A much earlier version of this chapter was presented at a conference at Dartmouth College. I thank the audience for their helpful feedback. I have also profited from detailed feedback from Chris Fuchs, Steven French, Michel Bitbol, and David Glick. As always, all remaining errors and omissions are entirely mine.